\documentclass{article}
\usepackage{epsfig}
\usepackage{graphicx}
\begin{document}

\title{\bf 
Effect of Atomic Coherence on Absorption in Four-level
Atomic Systems: an Analytical Study}
\author{S N Sandhya\\
Indian Institute of Technology \\
KANPUR 208016 INDIA\\
email: sns@iitk.ac.in}
\maketitle
\begin{abstract}
Absorption profile of a four-level ladder atomic system interacting 
with three driving fields is studied perturbatively and analytical results
are presented.
Numerical results where the driving field strengths are treated upto
all orders are presented. The absorption features
is studied in two regimes, i) the weak middle transition coupling, i.e.
 $\Omega_2 << \Omega_{1,3}$ and ii) the strong middle transition coupling
 $\Omega_2 >>\Omega_{1,3}$. In case i),
 it is shown that the ground state absorption and the saturation characteristics of the population of level 2 reveal deviation due to the presence
of upper level couplings. In particular, the saturation curve for the population of level 2
shows a dip 
for $\Omega_1 =  \Omega_3$. While the populations of levels 3 and 
4 show a 
maxima when this resonance condition is satisfied. Thus the resonance condition
provides a criterion 
for maximally populating the upper levels. 
A second order perturbation calculation reveals the nature of this 
minima (maxima). In the second case, I report two important features:
a) Filtering of the Aulter-Townes doublet in the three-peak absorption
profile of the ground state, which is achieved by detuning
 only the upper most coupling field, 
and b) control of line-width by controlling the strength of the upper coupling
fields. 
This filtering technique coupled with the control of linewidth
 could prove to be very useful 
for high resolution studies. 
\end{abstract}


\section{Introduction}
It is well known that the control of absorption and emission properties of 
atoms may be achieved by controlling the nature of atomic coherence. 
Driving field strengths and detunings provide the necessary tuning parameters
to achieve this control. Multi-level atoms have the advantage of providing
a larger set of tuning parameters. Three level atoms show a class of phenomena
which are absorption inhibitive \cite{scully} while four-level systems
have shown absorption inducive features at three photon resonance \cite{sandhya}.
The narrow absorption features which have been seen in four level systems
is attributed to the interaction of double
dark resonances \cite{ddk}. Scully et al \cite{Ye} report the occurrence of Doppler-free absorption, which has also been reported earlier \cite{sandhya}. 
Other features which have been reported in four-level systems include 
coherence switching \cite{ham}, photon switching (two-photon absorption)
\cite{yamomoto}, fast switching of nonlinear absorption \cite{yan2},
 enhancement and suppression of two photon absorption
\cite{harsha} and the occurrence of three peaked absorption \cite{ sandhya,
yan,yang,wei}.
In this paper we investigate the absorption properties and the modification
of the ground state absorption due to the additional upper transition
couplings.
In section II the details of the model is 
described and the
equation of motion is set up.
 Previous studies \cite{snsopt} have revealed that the dynamics of four-level
atoms interacting with three driving fields may be classified into two
broad domains, i) the weak middle transition coupling and ii) the strong
middle transition coupling. Hence, the analysis is presented separately for these two cases in sections III and IV respectively. 
In both the cases analytical results are derived perturbatively and
the qualitative features are compared with exact numerical calculations which are illustrated graphically.
\section{Equation of motion.}
The model we consider consists of a four-level ladder system shown in Fig 1.
The only dipole allowed transitions are $1\leftrightarrow2$, $2\leftrightarrow3$, and $3\leftrightarrow4$ coupled
respectively by the Rabi frequencies $\Omega_1,\Omega_2,\Omega_3$ and the Bohr
frequencies are denoted by $\omega_i,i=1,3$. The decay
constants corresponding to the respective levels are denoted by $\Gamma_i$,
i=1,4. The frequency of the applied fields are denoted by $\omega_{L_i},i=1,3$
 and the detunings are denoted by $\Delta_i,i=1,3$ corresponding to the three
couplings. The analysis presented here can also in principle be valid to any other type of
4-level system with transitions such that not more than two dipole
transitions share a single level. This could include 'N' type systems
as well as the mirror reflected 'N' type systems.
This particular ladder system may be identified,
for example with the Rb hyper fine levels. For instance, the $5s_{\frac{1}{2}},5p_{\frac{3}{2}}$ and $5d_{\frac{5}{2}}$ could correspond to the three
levels with an additional hyperfine level corresponding to either  $5p_{\frac{3}{2}}$ $ or$ $5d_{\frac{5}{2}}$. In fact the decay parameters of the model
have been chosen to be the same as the decay constants of these levels,
$\Gamma_2=6\gamma, \Gamma_3=\gamma$ and $\Gamma_4=\gamma$ in terms
of the ground state life time $\gamma$ which is chosen to be one for 
convenience. 
\begin{figure}
\centering
\includegraphics{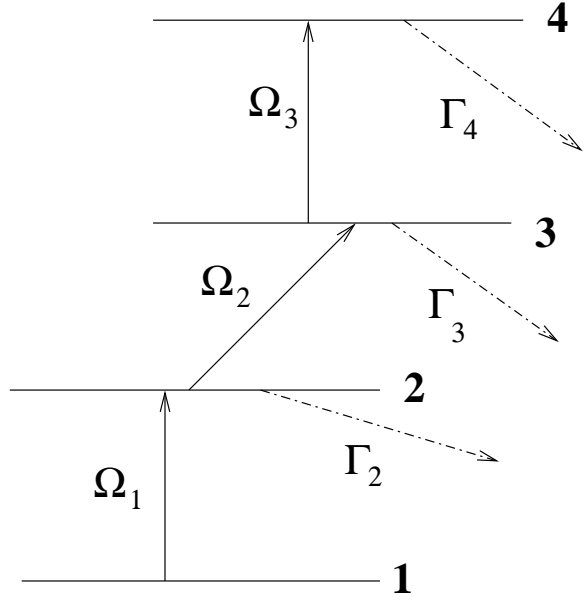}

\caption{ Four-level system interacting with three driving
fields of strengths $\Omega_1, \Omega_2,\Omega_3$ . $\Gamma_2,\Gamma_3,\Gamma_4$ are the decay constants of the corresponding levels.
}
\end{figure}
The density matrix equations (\cite{snsopt})  are written in the rotating wave approximation
as

\begin{eqnarray}
\frac {\partial {\rho}_{12}}{\partial t} &=& (-i \Delta_1 -\Gamma_{2}/2)
{\rho_{12}} -i \Omega_1 ({\rho_{22}}-{\rho_{11}}) + i \Omega_2
{\rho_{13}} \nonumber \\ 
\frac {\partial {\rho}_{23}}{\partial t} &=& (-i \Delta_2 -(\Gamma_{2}+\Gamma_3)/2)
{\rho_{23}} -i \Omega_1 {\rho_{13}}-i \Omega_2 ({\rho_{33}}-{\rho_{22}}) +
 i \Omega_3 {\rho_{24}} \nonumber \\
\frac {\partial {\rho}_{34}}{\partial t} &=& (-i \Delta_3 -(\Gamma_{3}+\Gamma_4)/2)
{\rho_{34}} -i \Omega_2 {\rho_{24}}-i \Omega_3 ({\rho_{44}}-{\rho_{33}}) 
 \nonumber \\
\frac {\partial {\rho}_{13}}{\partial t} &=& (-i( \Delta_1+\Delta_2) -\Gamma_{3}/2)
{\rho_{13}} -i \Omega_1 {\rho_{23}}+i \Omega_2 {\rho_{12}}+
i \Omega_3 {\rho_{14}}) \nonumber \\
\frac {\partial {\rho}_{14}}{\partial t} &=& (-i( \Delta_1+\Delta_2+\Delta_3) -\Gamma_{4}/2)
{\rho_{14}} -i \Omega_1 {\rho_{24}}+i \Omega_3 {\rho_{13}}
 \nonumber \\
\frac {\partial {\rho}_{24}}{\partial t} &=& (-i( \Delta_2+\Delta_3) -(\Gamma_{2}+\Gamma_4)/2)
{\rho_{24}} -i \Omega_1 {\rho_{14}}-i \Omega_2 {\rho_{34}}+
i \Omega_3 {\rho_{23}}) \nonumber \\
\frac {\partial {\rho}_{22}}{\partial t} &=&  -\Gamma_{2} {\rho_{22}}+
i \Omega_1 ({\rho_{21}}-{\rho_{12}}) + i \Omega_2
({\rho_{23}}-{\rho_{32}}) \nonumber \\
\frac {\partial {\rho}_{33}}{\partial t} &=&  -\Gamma_{3} {\rho_{33}}+
i \Omega_3 ({\rho_{34}}-{\rho_{43}}) - i \Omega_2
({\rho_{23}}-{\rho_{32}})  
 \nonumber \\
\frac {\partial {\rho}_{44}}{\partial t} &=&  -\Gamma_{3} {\rho_{44}}-
i \Omega_3 ({\rho_{34}}-{\rho_{43}})  
\label{spc}
\end{eqnarray}
where 
 $ \Gamma_i$ are the decay constants of the levels i and
$\Delta_i=\omega_{i,i+1}-\omega_i$ are the laser detunings
(${\rho_{ij}} = {\rho_{ji}}^{*}$ and ${\rm Tr}{\rho}=1).$
We study the steady state behaviour of the system for various values of the driving
field strengths and detunings. In the general situation, we present the
numerical solutions to the populations and coherences where the driving field strengths 
are treated up to all orders. 
The modification of the absorption profile
of the ground state as well as the upper levels will be studied analytically
by perturbatively solving for the density matrix elements in $\Omega_2$ or
$\Omega_1$ as the case may be. 
\section{Weak middle transition coupling.}
In the case when the transition
$2\leftrightarrow 3$ is weakly coupled, i.e. when $\Omega_2$ is very small compared to $\Omega_1,\Omega_3$, 
the equation of motion for the density matrix elements can be solved perturbatively in $\Omega_2$. The set of fifteen coupled equations
decouples into
three subsystems of equations which are solved independently to obtain
analytical solutions.
 In fact, the density matrix elements given by the sets 
$\{\rho_{11},\rho_{22},\rho_{12},\rho_{21} \},$$\{ \rho_{33},\rho_{34},\rho_{43},\rho_{44} \}$ and $\{ \rho_{23},\rho_{13}, 
\rho_{14},\rho_{24} \}$ form three subsystems.
The zeroth and first order solutions for 
$\rho_{ij},i,j=3,4$ remain zero, while the coherences $\{ \rho_{23},\rho_{24},
\rho_{13},\rho_{14} \} $ are nonzero in the first order and are listed in the Appendix.
 The steady state second order solutions for $\rho_{ij}$ are listed below:
\begin{eqnarray}
\rho_{22}^{(2)}=&(2 \bar \Gamma_2 \Omega_1^2(1-\rho_{33}^{(2)}-\rho_{44}^{(2)})
-\Omega_2 Im\rho_{23}^{(1)}(\Delta_1^2+\bar \Gamma_2^2)+\nonumber \\
&2 \Omega_1 \Omega_2(\Delta_1 Im\rho_{13}^{(1)}+ \bar \Gamma_2 Re\rho_{13}^{(1)})/{\cal D}_1\nonumber \\
\rho_{12}^{(2)}=&-i (\bar \Gamma_2 \Omega_1 (i \Delta_1-\bar \Gamma_2)(1-\rho_{33}^{(2)}-\rho_{44}^{(2)})+\bar \Gamma_2 \Omega_2 (i \Delta_1-\bar \Gamma_2)
 \rho_{13}^{(1)} \nonumber \\
&+2 \Omega_1 \Omega_2(i\Delta_1-\bar \Gamma_2) Im\rho_{23}^{(1)}-2\Omega_1^2 \Omega_2 (\rho_{13}^{(1)}-\rho_{31}^{(1)})/{\cal D}_1\nonumber \\
\rho_{33}^{(2)}=&(\bar \Gamma_4 \Omega_2 (\Delta_3^2+(\bar \Gamma_3+\bar \Gamma_4)^2) 
-2 \Omega_2 \Omega_3^2(\bar \Gamma_3+\bar \Gamma_4)) Im\rho_{23}^{(1)}\nonumber \\
&+2\Omega_2 \Omega_3\bar \Gamma_4(\Delta_3 Im\rho_{24}^{(1)}+(\bar \Gamma_3+\bar \Gamma_4)Re \rho_{24}^{(1)})/{\cal D}_3\nonumber \\
\rho_{44}^{(2)}=&(2 \Omega_2 \Omega_3\bar \Gamma_3 (\Delta_3 Im\rho_{24}^{(1)}+
(\bar \Gamma_3+\bar \Gamma_4)Re\rho_{24}^{(1)})
+\Omega_2\Omega_3^2(\bar \Gamma_3+\bar \Gamma_4)Im\rho_{23}^{(1)})/{\cal D}_3\nonumber \\
\rho_{34}^{(2)}=&i(2i \Omega_2 \Omega_3^2(\bar \Gamma_4+\bar \Gamma_3)Im\rho_{24}^{(1)}-\bar \Gamma_3 \bar \Gamma_4 \Omega_2(i \Delta_3-\bar \Gamma_3-\bar \Gamma_4)\rho_{24}^{(1)}\nonumber \\
&+\bar \Gamma_4\Omega_2 \Omega_3(i \Delta_3-(\bar \Gamma_3+\bar \Gamma_4))Im\rho_{23}^{(1)})/{\cal D}_3
\end{eqnarray} 

where ${\cal D}_1=\bar \Gamma_2(\Delta_1^2+\bar \Gamma_2^2+4 \Omega_1^2)$ and
${\cal D}_3=2 \Omega_3^2(\bar \Gamma_3+\bar \Gamma_4)^2-\bar \Gamma_3 \bar \Gamma_4(\Delta_3^2+
(\bar \Gamma_3+\bar \Gamma_4)^2)$.
Here, for convenience we have relabeled $ \Gamma_i/2$ everywhere
by $\bar \Gamma_i$ . The second order steady state solutions 
for the set of coherences $\{ \rho_{23},\rho_{13},\rho_{24},\rho_{14}\}$
are the same as the first order solutions given in the Appendix. 

Note that, in the limit $\Omega_2\rightarrow 0$ and $\Omega_3\rightarrow 0$,
the solution for $\rho_{12}$ and $\rho_{22}$ are given by the zeroth order solutions
which is the usual absorption and population of two-level system
driven by a field of strength $\Omega_1$. There is very little modification
in $\rho_{12}$ and $\rho_{22}$ due to the first order contribution.
However the inclusion of the second order contribution introduces 
additional terms. Before  investigating this further, let us look at the graphical
illustration of the ground state absorption as a function of detuning 
which has been solved numerically for all orders in the driving field
strengths. In Fig 2 the curve corresponding to (a) shows the absorption in the presence of only $1 \leftrightarrow 2$
coupling, (b) corresponds to the absorption of the ground state (g.s.) in the presence
of  the $1 \leftrightarrow 2$ as well as the $2 \leftrightarrow 3$ couplings
 and (c) shows the modification
of the g.s. absorption in  the presence of the $1 \leftrightarrow 2$ the first
excited state coupling $2 \leftrightarrow 3$ and the $3 \leftrightarrow 4$ coupling. Here the small
dip in the line center is not due to population trapping but due to the
population transfer to the upper levels which is because of the inclusion of the upper most transition coupling.
\begin{figure}
\centering
\includegraphics[width=10cm]{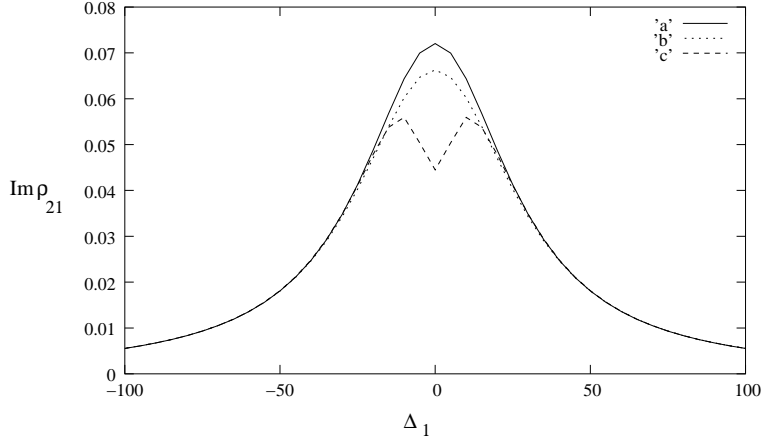}

\caption{$Im\rho_{21}$ as a function of the detuning $\Delta_1$.
a) $\Omega_1=20,\Omega_2=\Omega_3=0$,
b) $\Omega_1=20,\Omega_2=2,\Omega_3=0$, c)$\Omega_1=20,\Omega_2=2,\Omega_3=20$ and $\Gamma_2=6,\Gamma_3=\Gamma_4=1$. 
}
\end{figure}
This becomes more clear by looking at equations (2) and (3) which will be discussed shortly.
Let us look at the variation of the populations of 
various levels as a function of $\Omega_1$ and $\Omega_3$ for very
weak $\Omega_2$. Fig3-5 show the variation of populations w.r.t.
$\Omega_1$ and $\Omega_3$ which have been obtained numerically. Note that $\rho_{22}$ shows a dip in the absorption
when $\Omega_1=\Omega_3$. There is however a corresponding increase
in the populations of the upper levels. Fig4 and Fig5
show a sharp increase in the absorption by level 3 and 4 respectively  as $\Omega_1$ approaches $\Omega_3$
and shows a maxima for $\Omega_1=\Omega_3$ when
$\Omega_2=2$. 
\begin{figure}
\centering
\includegraphics[width=10cm]{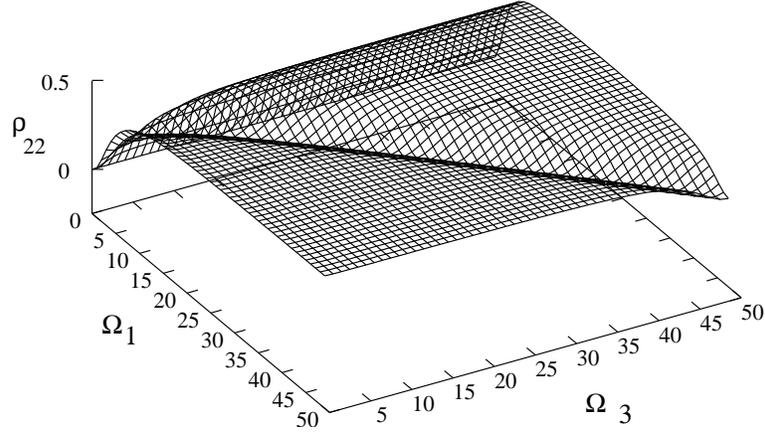}
\caption{$\rho_{22}$ as a function of the driving field strengths $\Omega_1$ and
$\Omega_3$ with $\Omega_2=2$ and $\Gamma_2=6,\Gamma_3=\Gamma_4=1$.
Occurence of a minima for $\Omega_1=\Omega_3$.
}
\end{figure}
\begin{figure}
\centering
\includegraphics[width=10cm]{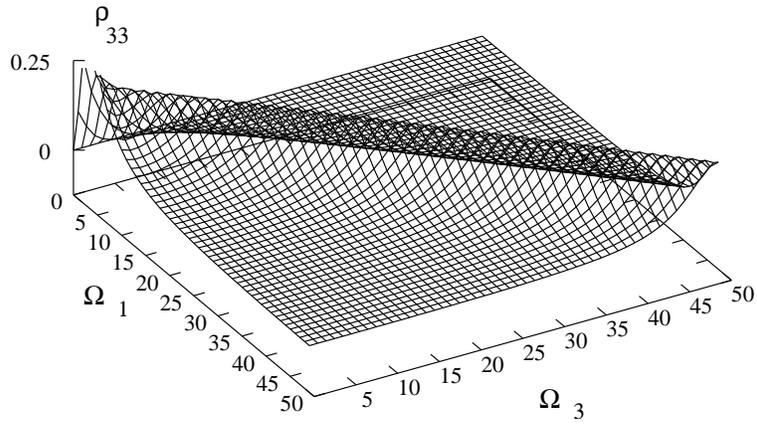}
\caption{$\rho_{33}$ as a function of the driving field strengths $\Omega_1$ and
$\Omega_3$ with $\Omega_2=2$.
Occurence of a  maxima for $\Omega_1=\Omega_3$.
}
\end{figure}
\begin{figure}
\centering
\includegraphics[width=10cm]{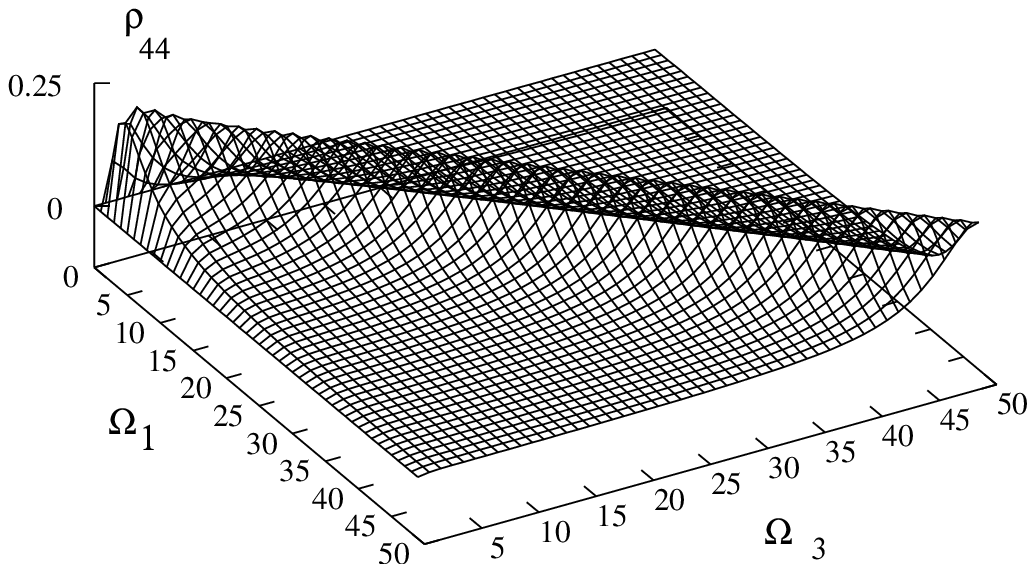}
\caption{$\rho_{44}$ as a function of the driving field strengths $\Omega_1$ and
$\Omega_3$ with $\Omega_2=2$ and $\Gamma_2=6,\Gamma_3=\Gamma_4=1$.
Occurence of a maxima for $\Omega_1=\Omega_3$.
}
\end{figure}

One can explain this analytically by looking at the second order solutions.
Consider
the situation when all the detunings are zero and ignoring the contribution
of the decay parameters $O(\Gamma_i^2)$  i.e. assuming $\Omega_{1,3}>>\Gamma_{2,3,4}$ for $\Delta_i=0$,
the populations  reduce to the simple form 

\begin{eqnarray}
\rho_{22}^{(2)} \approx&\rho_{22}^{(0)}+ \frac{\Omega_1^2 \Omega_2^2 \Omega_3^2( \bar \Gamma_3 \rho_{22}^{(0)}-\Omega_1 Im \rho_{21}^{(0)}  )
}{\bar \Gamma_4(\Omega_1^2-\Omega_3^2)+{\cal G}}\nonumber \\
Im\rho_{21}^{(2)}\approx & Im\rho_{21}^{(0)}+\frac{4 \Omega_2^2 \Omega_3^2 \Omega_1((\bar \Gamma_2 \rho_{22}^{(0)} 
(2 \bar \Gamma_3-\bar \Gamma_4)+Im\rho_{21}^{(0)}(-2 \bar \Gamma_2+\bar \Gamma_4)\Omega_1}
{\bar \Gamma_4(\Omega_1^2-\Omega_3^2)+{\cal G}}\nonumber \\
Im\rho_{34}^{(2)}\approx &\frac{\bar \Gamma_4 \Omega_2^2 \Omega_3(-\rho_{22}^{(0)}
 \bar \Gamma_3 +\Omega_1 Im \rho_{21}^{(0)})}{2 \bar \Gamma_4 \Omega_3^2(\Omega_1^2-\Omega_3^2)+{\cal G}}\nonumber \\
\rho_{33}^{(2)}\approx & \frac{\Omega_2^2(2\Omega_3^2(\bar \Gamma_4-\bar \Gamma_3) 
\rho_{22}^{(0)}+2 \Omega_3^2 \Omega_1 Im\rho_{21}^{(0)})}{2 \bar \Gamma_4 \Omega_3^2(\Omega_1^2-\Omega_3^2)+{\cal G}}\nonumber \\
\rho_{44}^{(2)}\approx& \frac{2 \Omega_2^2 \Omega_3^2(\bar \Gamma_3 \rho_{22}^{(0)}
-\Omega_1 Im\rho_{21}^{(0)})}{2 \bar \Gamma_4 \Omega_3^2(\Omega_1^2-\Omega_3^2)+{\cal G}}\nonumber \\
\end{eqnarray}

\begin{figure}
\centering
\includegraphics[width=10cm]{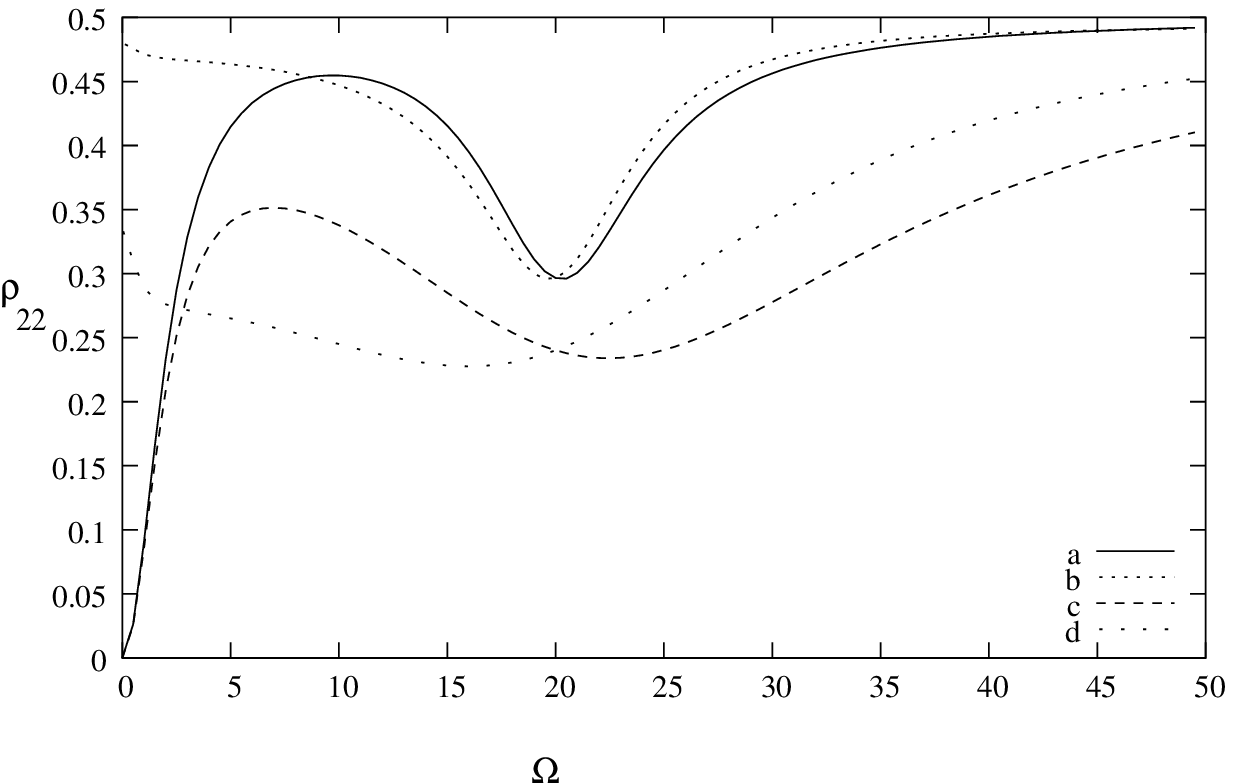}
\caption{$\rho_{22}$ as a function of the driving field strengths $\Omega$ 
a) $ \Omega_2=2, \Omega_1=20,\Omega=\Omega_3$ 
b) $ \Omega_3=20,\Omega_2=2,\Omega=\Omega_1$ 
c) $\Omega_2=8,\Omega_1=20,\Omega=\Omega_3$ 
d) $\Omega_3=20,\Omega_2=8,\Omega=
\Omega_1$. Here,
$\Gamma_2=6,\Gamma_3=\Gamma_4=1$.
}
\end{figure}
\begin{figure}
\centering
\includegraphics[width=10cm]{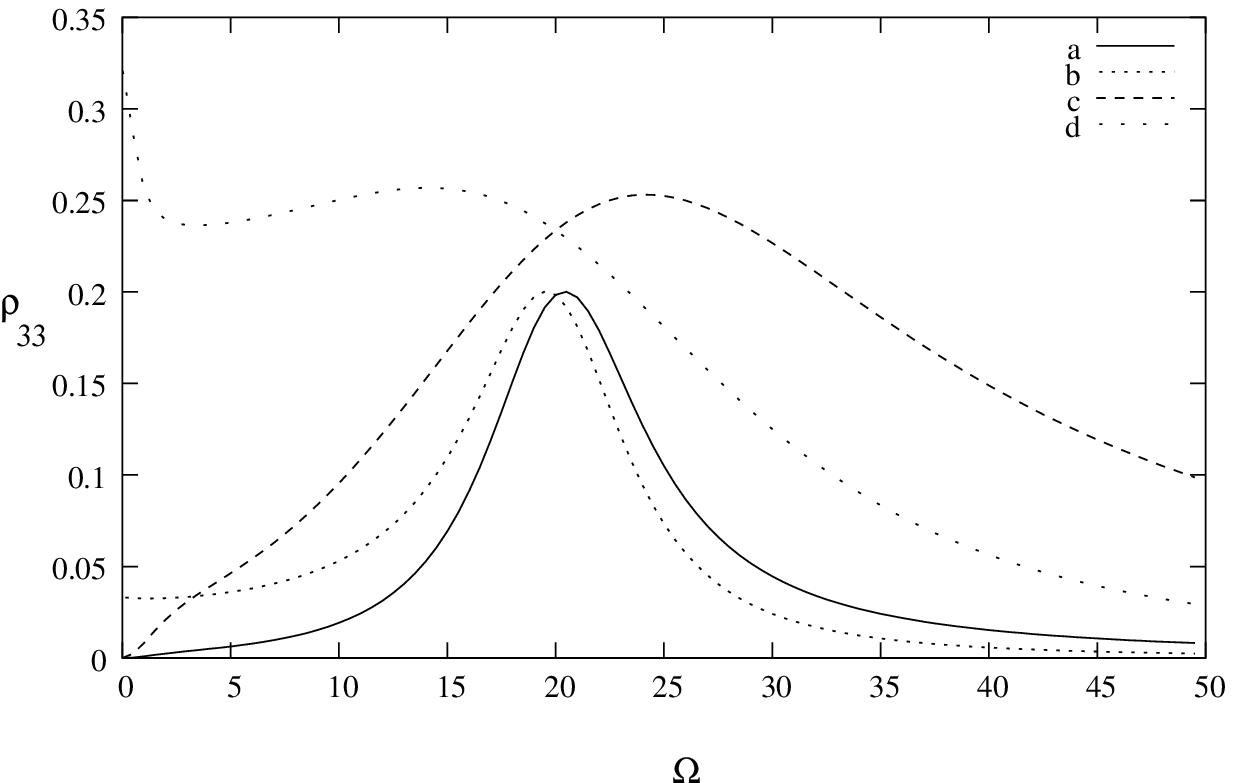}
\caption{$\rho_{33}$ as a function of the driving field strengths $\Omega$ 
a) $ \Omega_2=2, \Omega_1=20,\Omega=\Omega_3$ 
b) $ \Omega_3=20,\Omega_2=2,\Omega=\Omega_1$ 
c) $\Omega_2=8,\Omega_1=20,\Omega=\Omega_3$ 
d) $\Omega_3=20,\Omega_2=8,\Omega=
\Omega_1$. Here,
$\Gamma_2=6,\Gamma_3=\Gamma_4=1$.
}
\end{figure}
\begin{figure}
\centering
\includegraphics[width=10cm]{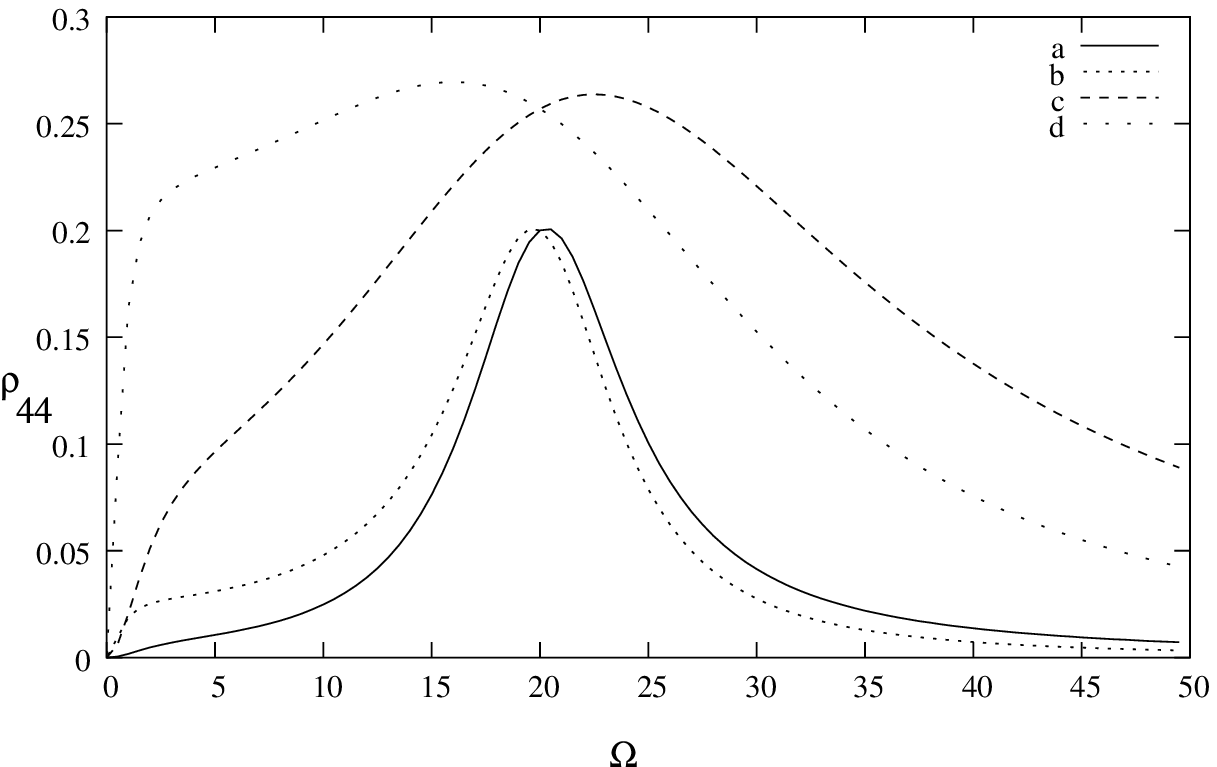}
\caption{$\rho_{44}$ as a function of the driving field strengths $\Omega_1$. 
a) $ \Omega_2=2, \Omega_1=20,\Omega=\Omega_3$ 
b) $ \Omega_3=20,\Omega_2=2,\Omega=\Omega_1$ 
c) $\Omega_2=8,\Omega_1=20,\Omega=\Omega_3$ 
d) $\Omega_3=20,\Omega_2=8,\Omega=
\Omega_1$. Here,
$\Gamma_2=6,\Gamma_3=\Gamma_4=1$.
}
\end{figure}
where $\rho_{22}^{(0)}=2\Omega_1^2/(\bar \Gamma_2^2+4\Omega_1^2)$ and $
Im \rho_{21}^{(0)}=\bar \Gamma_2 \Omega_1/(\bar \Gamma_2^2+\Omega_1^2)$ for
$\Delta_1=0$ and ${\cal G}$ includes terms of the $O(\bar \Gamma_i^2)$. It is obvious from these expressions that there is a 
resonance at $\Omega_1=\Omega_3$. The populations $\rho_{33}^{(2)}$
and $\rho_{44}^{(2)}$ show a sharp rise as $\Omega_1\rightarrow \Omega_3$
since the numerator is positive. Thus to populate the levels 3 and 4
maximally when the middle level coupling is weak, it is sufficient
to match the requirement $\Omega_1=\Omega_3$. The absorption profile
$Im \rho_{34}^{(2)}$ also shows a similar trend. While the second 
term in $\rho_{22}^{(2)}$
and $\rho_{21}^{(2)}$  are negative and the role of the resonance in the denominator
is to introduce a minima when $\Omega_1=\Omega_3$. 
Thus the role of the second order term 
in this case, is to  introduce a dip in the line center in an otherwise
two-level absorption profile and hence a depletion in the population of level 2. As one increases $\Omega_2$ this perturbative
expansion is no longer valid. Moreover, the numerical calculations
upto all orders in $\Omega_2$ reveal the fact that the resonance condition
becomes weaker and the sharp features vanish. This is demonstrated
graphically in Fig6-8. 
 
Thus, in a four-level system with middle transition coupling very weak, the condition for maximally populating the upper levels is to satisfy this
resonance condition. The population transfer to higher levels occurs at
the cost of depletion of population in the second level. This modification
is mediated by the contribution of the coherences, which at three photon
resonance are given by 
$\rho_{23}^{(1)}\approx \Omega_2 Im\rho_{21}^{(0)}/
{\cal G}_0,
\rho_{13}^{(1)}\approx \Omega_1 \Omega_2 \rho_{22}^{(0)}/{\cal G}_0,
\rho_{24}^{(1)}\approx \Omega_2 \Omega_3 \rho_{22}^{(0)}/{\cal G}_0$
where ${\cal G}_0=\bar \Gamma_4(\Omega_1^2-\Omega_3^2).$

\section{Strong middle transition coupling}
I next consider the case when the $2\rightarrow 3$ transition is strongly coupled i.e.,
$\Omega_2>>\Omega_1,\Omega_3$.
Treating $\Omega_1$ perturbatively, the first order absorption of the
ground state takes the simple form \cite{sandhya}
\begin{equation}
Im\rho_{21}^{(1)}=\frac{L_{12}+\Omega_3^2/L_{123}\Omega_1}{
(\Omega_2^2+L_1 L_{12}+\Omega_3^2 L_1/L_{123})}
\end{equation}
where $L_1=(-i \Delta_1-\Gamma_2/2),L_{12}=(-i(\Delta_2+\Delta_1)-\Gamma_3/2)
$ and $L_{123}= (-i(\Delta_1+\Delta_2+\Delta_3)-\Gamma_4/2)$. We now look
at some simple situations, for example when $\Omega_2\rightarrow 0,\Omega_3\rightarrow 0$, the 
absorption becomes proportional to $\Gamma_2 \Omega_1/(\Delta_1^2+\Gamma_2^2)$
which is the ground state absorption for a weak excitation and observe that
the width is proportional to $\Gamma_2$. Next, consider only $\Omega_3\rightarrow 0$.
The absorption is now given by
\begin{equation}
Im\rho_{21}=\frac{\Omega_1(\Delta_1^2 \bar \Gamma_2 +\bar \Gamma_3(\bar \Gamma_3 \bar \Gamma_2
+\Omega_2^2))}{\Delta_1^4+\Delta_1^2(\bar \Gamma_2^2+\bar \Gamma_3^2-2 \Omega_2^2)+
(\bar \Gamma_2\bar \Gamma_3+\Omega_2^2)^2}
\end{equation}
where $\Delta_2$ is assumed to be zero.
The denominator is quadratic in $\Delta_1^2$ and hence has two peaks corresponding
to the usual Autler-Townes doublet. The width of this EIT window is proportional to
$\sqrt(\bar \Gamma_2^2+\bar \Gamma_3^2-2 \Omega_2^2)^2-4(\bar \Gamma_2\bar \Gamma_3+\Omega_2^2)$.
The absorption for $\Delta_1=0$ (at the line center) is proportional
to $\Omega_1 \bar \Gamma_3/(\bar \Gamma_2\bar \Gamma_3+\Omega_2^2)$ which is negligible
since it is of the order $\Omega_1/\Omega_2^2$.
The absorption in the presence of the third driving field for $\Delta_2=0,
\Delta_3=0$ takes the
simplified form
\begin{equation}
Im\rho_{21}^{(1)}=\frac{\Omega_1(\Delta_1^2(\Delta_1^2\bar \Gamma_2+\bar \Gamma_2(\bar \Gamma_3^2+\bar \Gamma_4^2)
+\bar \Gamma_3 \Omega_2^2-2 \bar \Gamma_2 \Omega_3^2)+
T_1)}
{\Delta_1^2(\Delta_1^2-\Gamma_p-
\Omega_2^2-\Omega_3^2)^2+(\Delta_1^2\Gamma_s-
\bar \Gamma_2\bar \Gamma_3\bar \Gamma_4 -\bar \Gamma_4\Omega_2^2-\bar \Gamma_2\Omega_3^2)^2}
\end{equation}
where $\Gamma_s=\bar \Gamma_2+\bar \Gamma_3+\bar \Gamma_4$, $\Gamma_p=\bar \Gamma_2\bar \Gamma_3+
\bar \Gamma_3\bar \Gamma_4+\bar \Gamma_4\bar \Gamma_2$ and 
$T_1=\bar \Gamma_2\bar \Gamma_3^2\bar \Gamma_4^2
+\bar \Gamma_3\bar \Gamma_4^2 \Omega_2^2+2 \bar \Gamma_2\bar \Gamma_3 \bar \Gamma_4
\Omega_3^2+\bar \Gamma_4 \Omega_3^2\Omega_2^2+\bar \Gamma_2 \Omega_3^4)$.
The denominator is a cubic equation in $\Delta_1^2$ with the three roots
corresponding to the three 
absorption peaks, two of them correspond
to the usual Autler -Townes doublet and the absorption at the line center
is due to the three photon interaction and is proportional to
$\Omega_1 \Omega_3^2/(\bar \Gamma_4 \Omega_2^2+\bar \Gamma_2 \Omega_3^2)$ which occurs when
all the detunings are zero. This three peaked absorption in four level systems has been observed earlier and has been frequently referred 
to as the splitting of the EIT window. While the role of the strong
$2\leftrightarrow 3$ coupling was to introduce transparency at the line center, the role of the $3 \leftrightarrow 4$ coupling
is to induce a narrow absorption at the line center within the
EIT window. Thus the two photon absorption was responsible for the coherence $\rho_{31}$ which in turn lead to non-absorption while, the three photon absorption which introduced a coherence between levels
1 and 4 ($\rho_{41}$) which in turn induced this sharp absorption feature. Thus there is a contrast in the nature of the quantum interference in the
case of two and three photon interaction. 

On the other hand consider the case when the
detuning $\Delta_3 $ is nonzero. Let $\Delta_1=0,\Delta_2=0$, the absorption 
now reduces to
\begin{equation}
Im\rho_{21}^{(1)}= \frac{\Omega_1(\Delta_3^2\bar \Gamma_3(\bar \Gamma_2\bar \Gamma_3+
\Omega_2^2)+(\bar \Gamma_3\bar \Gamma_4+\Omega_3^2)(\bar \Gamma_4\Omega_2^2+\bar \Gamma_2
(\bar \Gamma_3\bar \Gamma_4+\Omega_3^2)))}{\Delta_3^2(\bar \Gamma_2\bar \Gamma_3+\Omega_2^2)^2+
(\bar \Gamma_4\Omega_2^2+\bar \Gamma_2(\bar \Gamma_3\bar \Gamma_4+\Omega_3^2))^2}
\end{equation}
This is a Lorentzian with width proportional to $\bar \Gamma_2 \Omega_3^2/\Omega_2^2$. 
This implies that the line width can be subnatural ($<\Gamma_2$) for $\Omega_3<\Omega_2$. Notice the novel feature here which is the
absence of the Autler-Townes doublet unlike the case when only $\Delta_1$
was nonzero. Thus, we now have obtained a powerful technique for filtering
the Autler-Townes doublet and retaining only the narrow absorption at the
line center in addition to gaining control over the linewidth. 
This could prove to be a useful technique in  high resolution spectroscopy.

A numerical illustration of these novel features is presented  
in Fig 9 where the steady state solutions are obtained for all
orders in the driving field strengths.
\begin{figure}
\includegraphics[totalheight=1.5in]{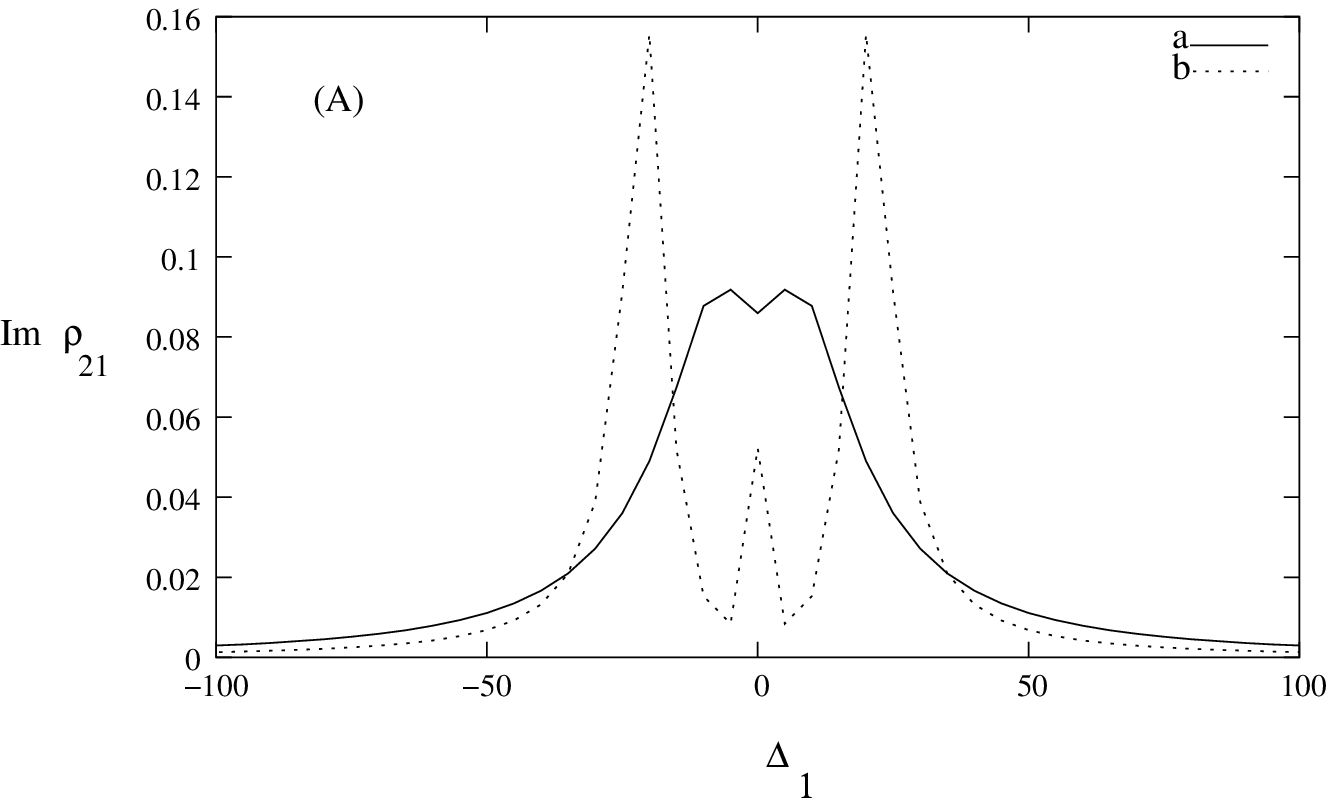}
\includegraphics[totalheight=1.5in]{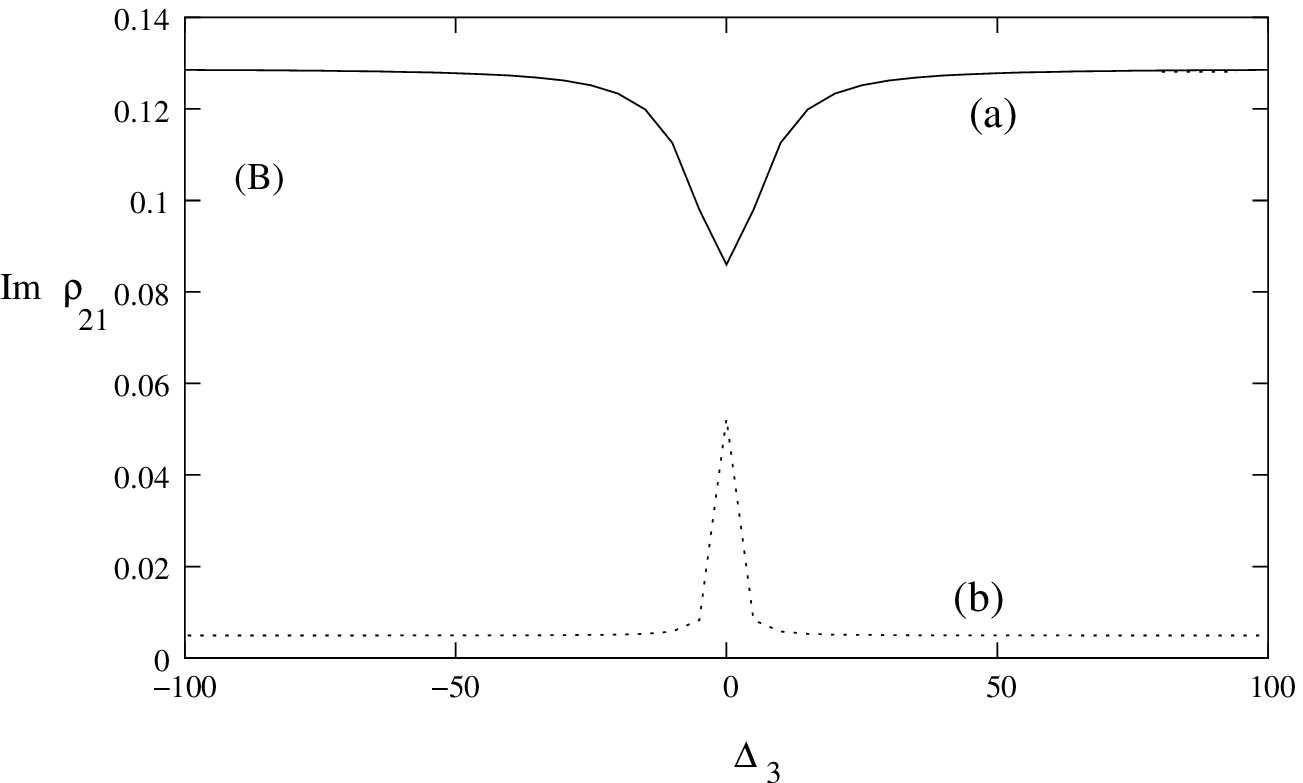}\\
\includegraphics[totalheight=1.5in]{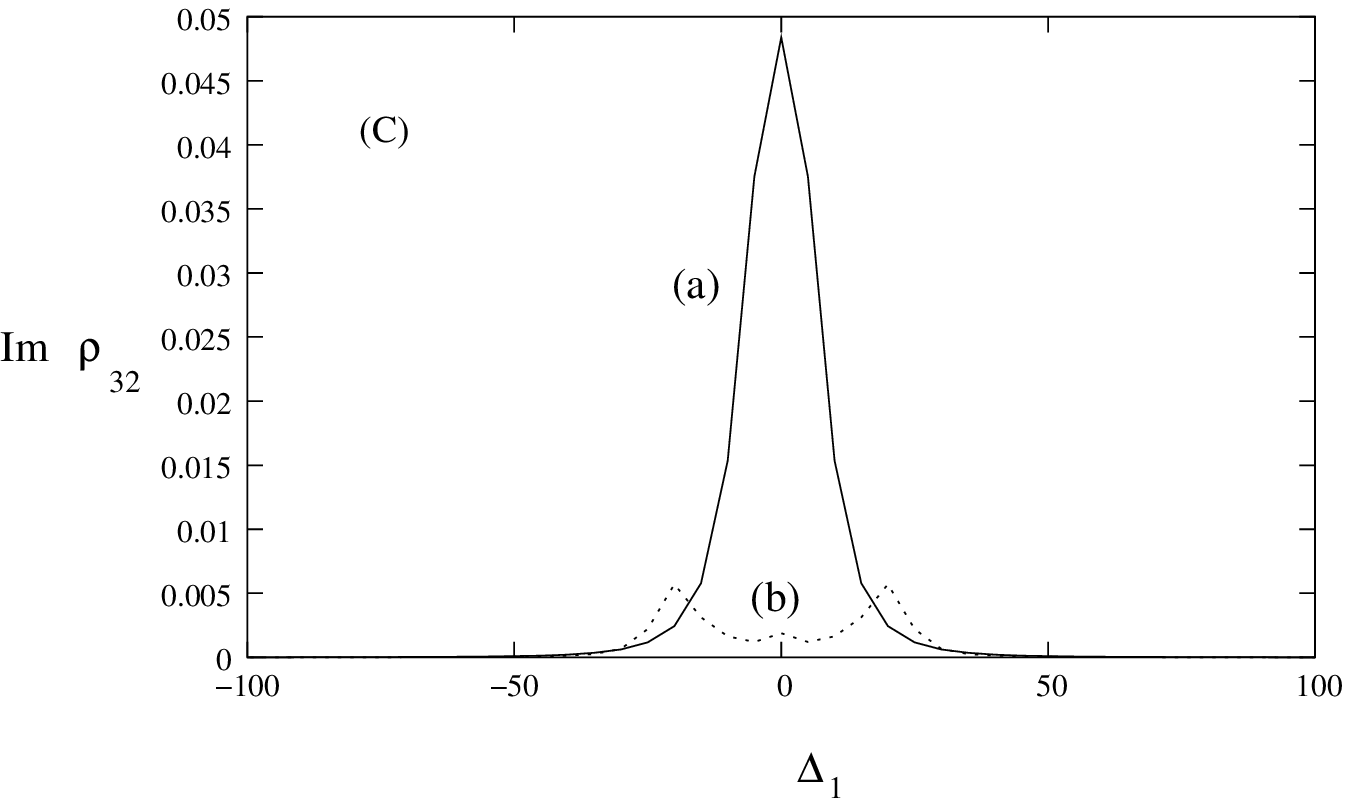}
\includegraphics[totalheight=1.5in]{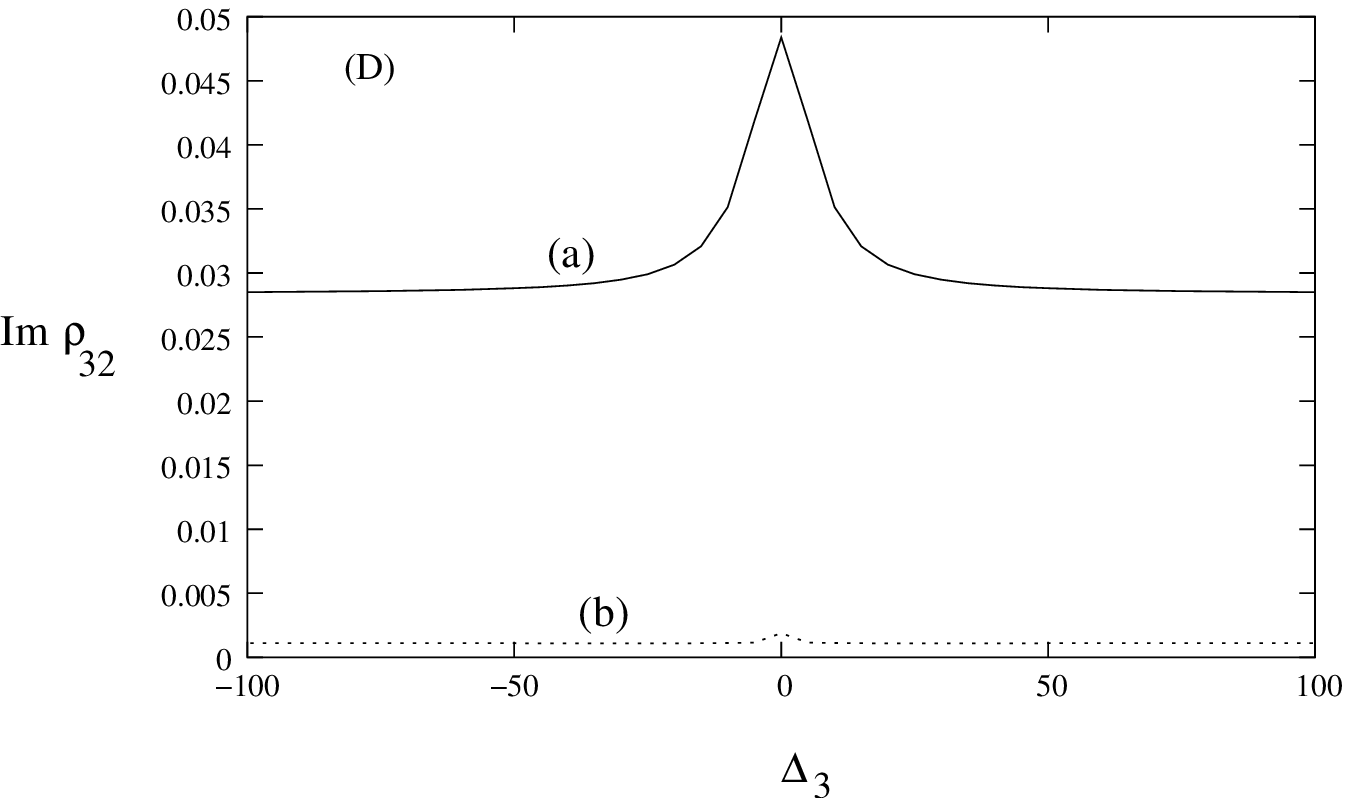}\\
\includegraphics[totalheight=1.5in]{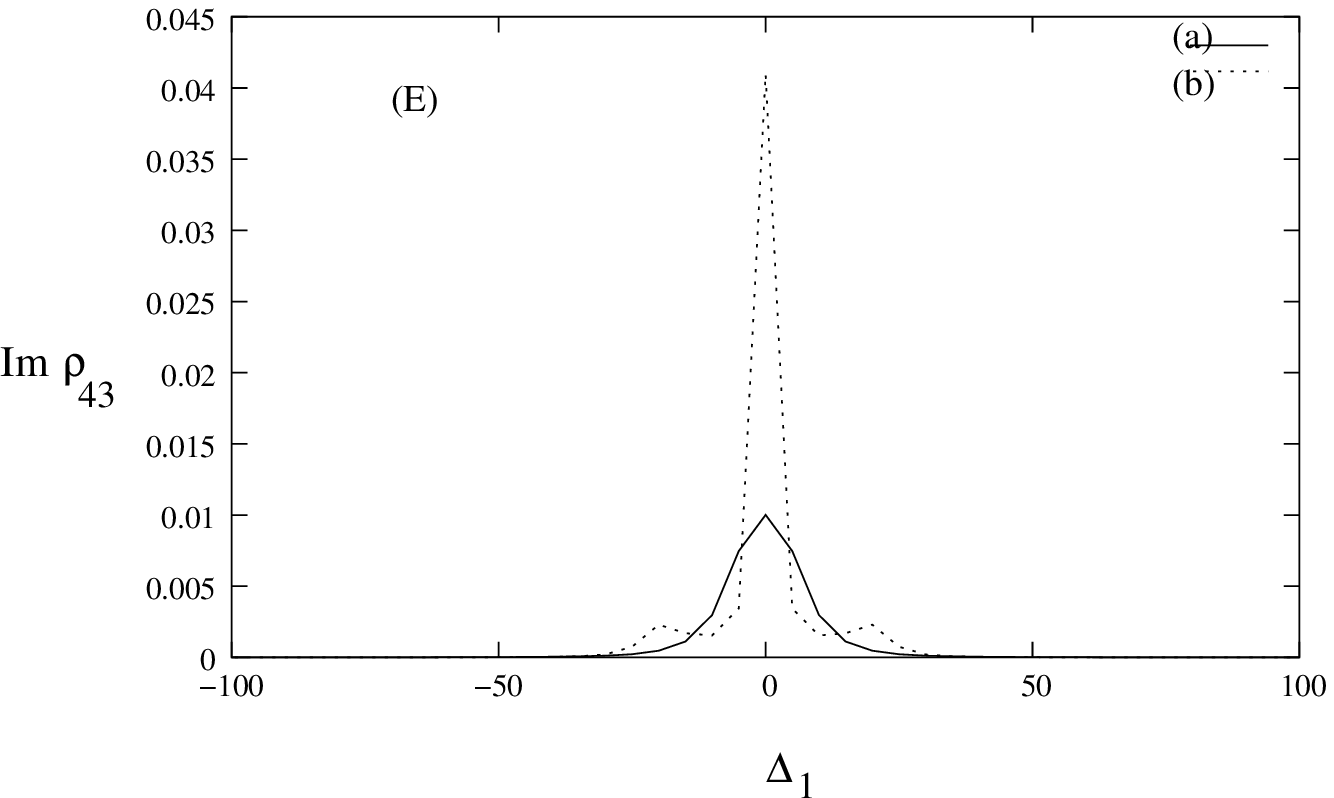}
\includegraphics[totalheight=1.5in]{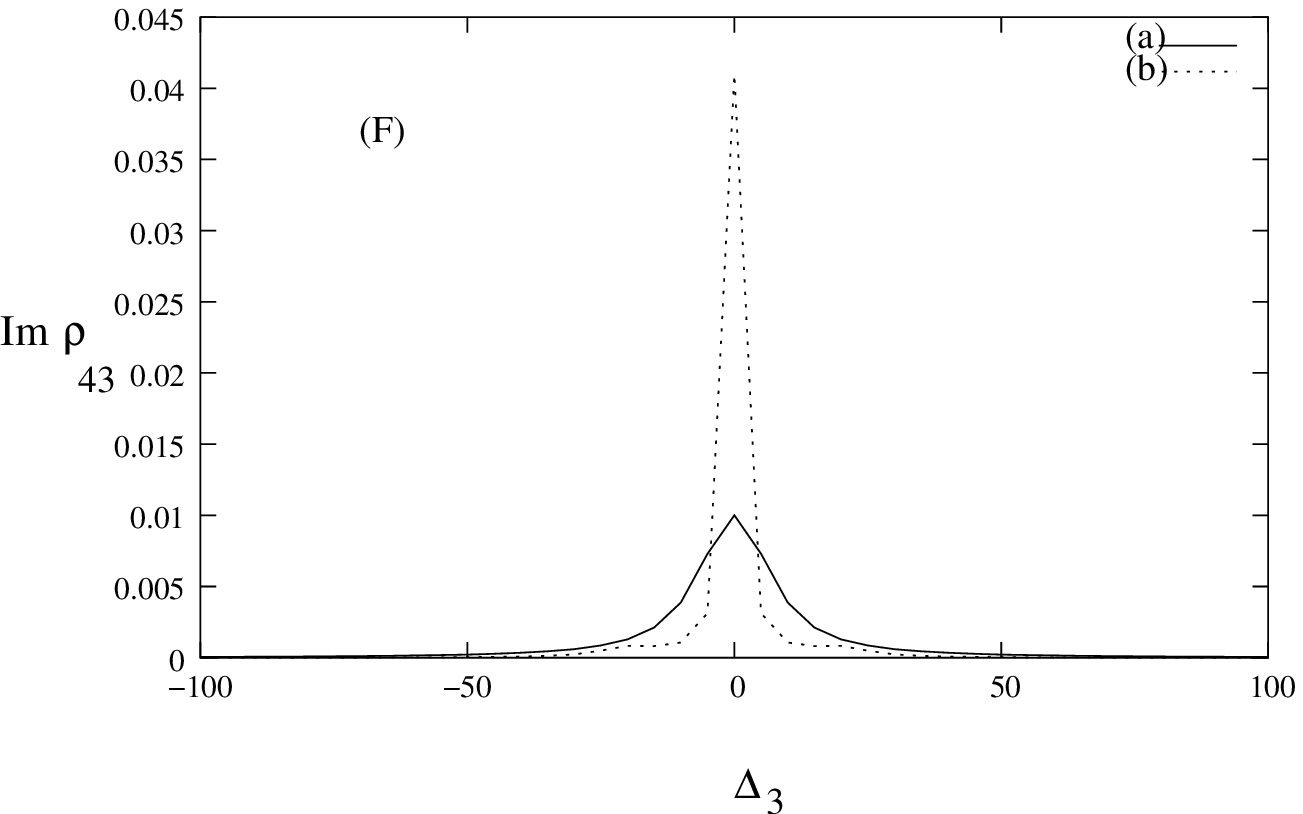}\\
\caption{The variation of absorption for (a)$\Omega_2=2,\Omega_1=\Omega_3=20$ and
(b)  $\Omega_2=20,\Omega_1=\Omega_3=4$ .A)$Im\rho_{21}$ with $\Delta_1$ B)$Im\rho_{21}$
with $\Delta_3$.C)$Im\rho_{32}$ with $\Delta_1$
D)$Im\rho_{32}$ with $\Delta_3$
E)$Im\rho_{43}$ with $\Delta_1$
F)$Im\rho_{43}$ with $\Delta_3$ with the $\Gamma_i,i=2,4$ the same
as in the previous figures.}
\end{figure}
Figure 9 clearly shows the difference in the behaviour depending
on the strength of the middle transition coupling. a) corresponds to weak $\Omega_2$
while b) corresponds to strong $\Omega_2$. Fig 9A-b shows the
three peaked absorption while Fig9B-b shows the elimination of the 
Autler-Townes doublet by detuning the driving field coupling 3-4 transition
as predicted by the perturbative calculations.
Likewise for small $\Omega_2$, only the narrow absorption dip at the
line center is seen by detuning $\Delta_3$(Fig 9A-a and Fig9B-a).
Again, $Im \rho_{32}$ for
small $\Omega_2$ is very strong (Fig9C-a) as the
perturbation calculation indicate (section III) and this absorption is not very 
significant for large $\Omega_2$ since the resonance condition weakens and 
the contribution of higher order terms in $\Omega_2$ dominate.
Lastly, the absorption profile of the uppermost transition does
not seem to vary much with either of the detunings $\Delta_1$
or $\Delta_3$ (Fig9-Eb,Fig9F-b). This is very much similar to the narrow
g.s. absorption at the line center. Further, both the absorption of the
g.s.(line center) and the upper most transition $Im\rho_{43}$ are unaffected
by the detunings $\Delta_1/\Delta_3$ since they occur at exact three photon
resonance. One would expect this absorption feature also to be Doppler-free
like the g.s. absorption \cite{sandhya}. 
\begin{figure}
\centering
\includegraphics[width=10cm]{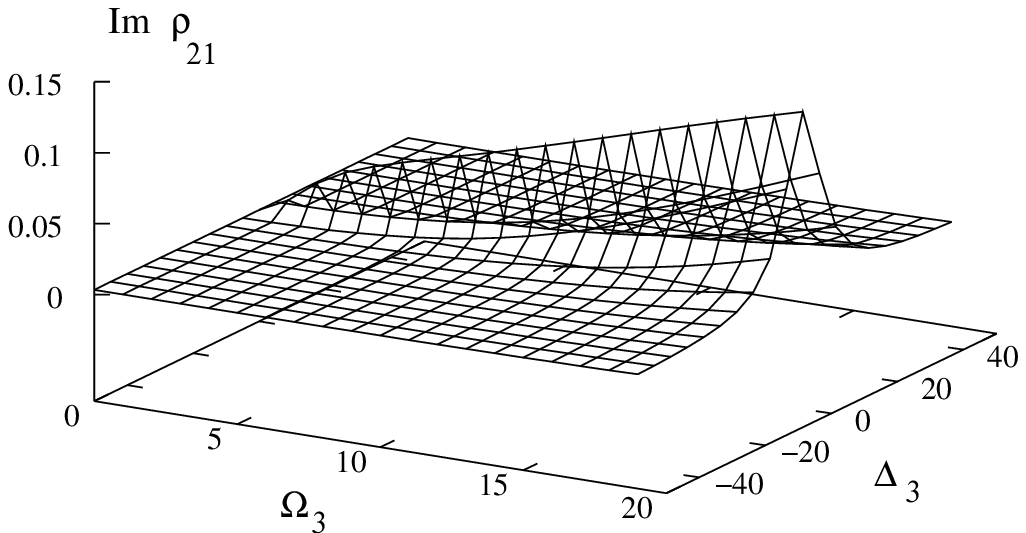}
\caption{$Im\rho_{21}$ as a function 
of $\Omega_3$ and $\Delta_3$ with parameter values 
$\Delta_1=\Delta_2=0$ 
$\Omega_1=4,\Omega_2=20$ and
$\Gamma_i,i=2,4$ the same as the previous figures.
}
\end{figure}
\begin{figure}
\centering
\includegraphics[width=10cm]{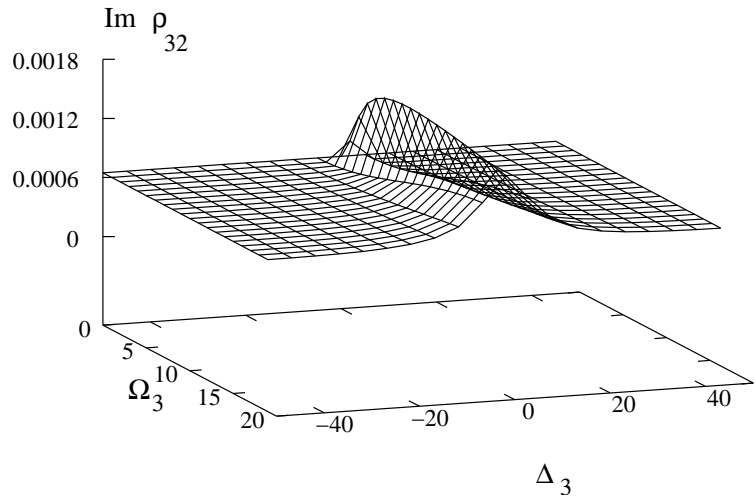}
\caption{$Im\rho_{32}$ as a function 
of $\Omega_3$ and $\Delta_3$ 
with $\Delta_1=\Delta_2=0$ and
the rest of the parameters 
the same as in the previous figure.
}
\end{figure}
\begin{figure}
\centering
\includegraphics[width=10cm]{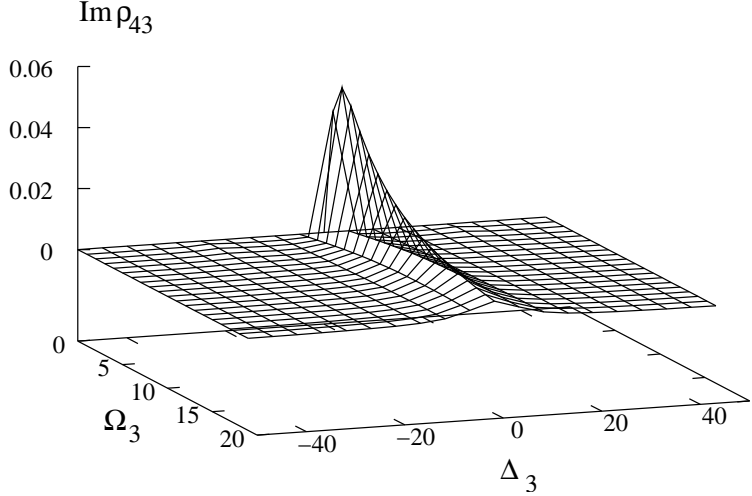}
\caption{$Im\rho_{43}$ as a function 
of $\Omega_3$ and $\Delta_3$ for parameter values the same
as in previous figure. 
}
\end{figure}
 
Next the variation of the absorption of the g.s.,
$Im \rho_{21}$, the first excited state $Im \rho_{32}$ and the
second excited state $Im \rho_{43}$ with the driving field strength
$\Omega_3$ and the detuning $\Delta_3$ are numerically illustrated
in Fig10-12. The purpose here, is to demonstrate the three photon
effect. While the ground state absorption increases with $\Omega_3$,
the middle and the upper transition absorptions rise sharply for small
$\Omega_3$ and fall rapidly to zero. All the absorption occurs only for $\Delta_3=0$. The variation with $\Delta_2$ shows lot more
structure and will be more appropriate to present it else where.

To summarise, it is shown that by satisfying the resonance condition
$\Omega_1=\Omega_3$ one can maximally populate the upper levels at the
cost of depleting the population of level 2 (for small $\Omega_2$).
In the case of large $\Omega_2$ it is shown that one can
get narrow absorption features at the line center by filtering
of the Autler-Townes doublet. One also has a control over the line-width
by controlling $\Omega_2$ and $\Omega_3$. The qualitative features predicted
by perturbative calculations seem to agree very well with the numerical
results.
\section{Appendix}
The zeroth order equation in $\Omega_2$ are given by
\begin{eqnarray}
\frac {\partial {\rho^{(0)}}_{12}}{\partial t} &=& (-i \Delta_1 -\Gamma_{2}/2)
{\rho^{(0)}_{12}} -i \Omega_1 ({\rho^{(0)}_{22}}-{\rho^{(0)}_{11}}) 
\nonumber \\ 
\frac {\partial {\rho^{(0)}}_{22}}{\partial t} &=&  -\Gamma_{2} {\rho^{(0)}_{22}}+
i \Omega_1 ({\rho^{(0)}_{21}}-{\rho^{(0)}_{12}}) 
\nonumber \\
\frac {\partial {\rho^{(0)}}_{21}}{\partial t} &=& (i \Delta_1 -\Gamma_{2}/2)
{\rho^{(0)}_{21}} +i \Omega_1 ({\rho^{(0)}_{22}}-{\rho^{(0)}_{11}})\nonumber 
\end{eqnarray}
the rest of the density matrix elements are zero (steady state).
The first order  equation do not contribute to the solutions 
of $\rho_{ij},i,j=1,2$. The density matrix elements $\rho_{ij},i,j=3,4$ remain zero. Hence we furnish below the first order equations
for the coherences only.
\begin{eqnarray}
\frac {\partial {\rho}_{23}^{(1)}}{\partial t} &=& (-i \Delta_2 -(\Gamma_{2}+\Gamma_3)/2)
{\rho_{23}} -i \Omega_1 {\rho_{13}^{(1)}}-i \Omega_2 (-{\rho_{22}^{(0)}}) +
 i \Omega_3 {\rho_{24}^{(1)}} \nonumber \\
\frac {\partial {\rho}_{13}^{(1)}}{\partial t} &=& (-i( \Delta_1+\Delta_2) -\Gamma_{3}/2)
{\rho_{13}^{(1)}} -i \Omega_1 {\rho_{23}^{(1)}}+i \Omega_2 {\rho_{12}^{(0)}}+
i \Omega_3 {\rho_{14}^{(1)}}) \nonumber \\
\frac {\partial {\rho}_{14}^{(1)}}{\partial t} &=& (-i( \Delta_1+\Delta_2+\Delta_3) -\Gamma_{4}/2)
{\rho_{14}^{(1)}} -i \Omega_1 {\rho_{24}^{(1)}}+i \Omega_3 {\rho_{13}^{(1)}}
 \nonumber \\
\frac {\partial {\rho}_{24}^{(1)}}{\partial t} &=& (-i( \Delta_2+\Delta_3) -(\Gamma_{2}+\Gamma_4)/2)
{\rho_{24}^{(1)}} -i \Omega_1 {\rho_{14}^{(1)}}+
i \Omega_3 {\rho_{23}^{(1)}}) \nonumber \\
\end{eqnarray}
the first order steady state solutions of these equations are given by
\begin{eqnarray}
\rho_{23}^{(1)}=&\Omega_2(\Omega_1\rho_{12}^{(0)}(\Omega_1^2-\Omega_3^2+L_{123}L_{23})+\nonumber \\
&i \rho_{22}^{(0)}(L_{12}L_{23}L_{123}+L_{12}\Omega_1^2+L_{23}\Omega_3^2))/{\cal D}_2\nonumber \\
\rho_{13}^{(1)}=&\Omega_2(i \rho_{12}^{(1)}(L_{123}L_{2}L_{23}+L_2 \Omega_1^2+
L_{123}\Omega_3^2)-\nonumber \\
&\Omega_1 \rho_{22}^{(1)}(\Omega_1^2-\Omega_3^2+L_{123}L_{23}))/{\cal D}_2 \nonumber \\
\rho_{24}^{(1)}=&- \Omega_2\Omega_3(\rho_{22}^{(0)}(L_{12}L_{123}-\Omega_1^2
+\Omega_3^2)+\nonumber \\
&i \Omega_1\rho_{12}^{(0)}(L_{123}+L_{2}))/{\cal D}_2 \nonumber\\
\rho_{14}^{(1)}=& \Omega_2\Omega_3(\rho_{22}^{(0)}(\Omega_1^2-\Omega_3^2-
L_{12}L_{123})+\nonumber \\
&i\Omega_1 \rho_{12}^{(0)}(L_{123}+L_{2}))/{\cal D}_2 \nonumber
\end{eqnarray}
where ${\cal D}_2=((\Omega_1^2-\Omega_3^2)^2 +\Omega_1^2(L_{123}L_{23}+L_{12}L_2)+\Omega_3^2(L_{12}L_{123}+L_{2}L_{23})+L_{12}L_{23}L_{123}L_{2})$ and
$L_2=(-i\Delta_2-(\Gamma_2+\Gamma_3)/2),L_{23}=(-i(\Delta_2+\Delta_3)-(\Gamma_2+\Gamma_4)/2)$ while $L_{12}$ and $L_{123}$ are given in the text.
The second order equations are given by
\begin{eqnarray}
\frac {\partial {\rho}_{12}^{(2)}}{\partial t} &=& (-i \Delta_1 -\Gamma_{2}/2)
{\rho_{12}^{(2)}} -i \Omega_1 ({\rho_{22}^{(2)}}-{\rho_{11}^{(2)}}) + i \Omega_2
{\rho_{13}^{(1)}} \nonumber \\ 
\frac {\partial {\rho}_{22}^{(2)}}{\partial t} &=&  -\Gamma_{2} {\rho_{22}^{(2)}}+
i \Omega_1 ({\rho_{21}^{(2)}}-{\rho_{12}^{(2)}}) + i \Omega_2
({\rho_{23}^{(1)}}-{\rho_{32}^{(1)}})\nonumber \\ 
\frac {\partial {\rho}_{34}^{(2)}}{\partial t} &=& (-i \Delta_3 -(\Gamma_{3}+\Gamma_4)/2)
{\rho_{34}^{(2)}} -i \Omega_2 {\rho_{24}^{(1)}}-i \Omega_3 ({\rho_{44}^{(2)}}-{\rho_{33}^{(2)}}) 
 \nonumber \\
\frac {\partial {\rho}_{33}^{(2)}}{\partial t} &=&  -\Gamma_{3} {\rho_{33}^{(2)}}+
i \Omega_3 ({\rho_{34}^{(2)}}-{\rho_{43}^{(2)}}) - i \Omega_2
({\rho_{23}^{(1)}}-{\rho_{32}^{(1)}})\nonumber \\  
\frac {\partial {\rho}_{44}^{(2)}}{\partial t} &=&  -\Gamma_{3} {\rho_{44}^{(2)}}-
i \Omega_3 ({\rho_{34}^{(2)}}-{\rho_{43}^{(2)}}) \nonumber 
\label{spc}
\end{eqnarray}
So far, we dealt with equations where $\Omega_2$ was treated 
perturbatively. I furnish below the relevant first order equations where
$\Omega_1$ is treated perturbatively,
\begin{eqnarray}
\frac {\partial {\rho}_{12}^{(1)}}{\partial t} &=& (-i \Delta_1 -\Gamma_{2}/2)
{\rho_{12}^{(1)}} -i \Omega_1  + i \Omega_2
{\rho_{13}^{(1)}} \nonumber \\ 
\frac {\partial {\rho}_{13}^{(1)}}{\partial t} &=& (-i( \Delta_1+\Delta_2) -\Gamma_{3}/2)
{\rho_{13}^{(1)}} +i \Omega_2 {\rho_{12}^{(1)}}+
i \Omega_3 {\rho_{14}^{(1)}}) \nonumber \\
\frac {\partial {\rho}_{14}^{(1)}}{\partial t} &=& (-i( \Delta_1+\Delta_2+\Delta_3) -\Gamma_{4}/2)
{\rho_{14}^{(1)}} +i \Omega_3 {\rho_{13}^{(1)}}\nonumber
\end{eqnarray}
\section*{Acknowledgments}
I thank Professor G S Agarwal for useful discussions. I wish to thank the Department of Science and Technology for Financial
support under the WOS-A scheme.

\end{document}